\documentclass{ws-p8-50x6-00}

\begin{document}

\title{The underlying event  in jet and minimum bias events 
at the Tevatron}

\author{Valeria Tano\\
for the CDF collaboration}

\address{Max-Planck-Institut f\"ur Physik,
F\"ohringer Ring, 6 
80333 Munich, Germany \\
E-mail: tano@mppmu.mpg.de}


\maketitle

\abstracts{We describe a study of the \it{underlying event}\rm~ in jet and minimum bias events using data from the CDF detector. The \it{underlying event}\rm~ contribution to the jet energy has been calculated in jet events and compared to the results of two Monte Carlo programs: Herwig and Pythia. The analysis has been carried out at two different center of mass energies: $\sqrt{s}=$1800 and 630 GeV. For most observables, good agreement is obtained with at least one of the Monte Carlo programs. Neither program describes all features of the data.}

{\section{A measure of the underlying event in jet events}

In hadron-hadron collisions, in addition to the  hard interaction that produces the jets in the final state, there is also an \it{underlying event}\rm, originating mostly from soft spectator interactions. 
In some of the events, the spectator interaction may be hard enough to produce soft jets (\it{double parton scattering}\rm)~\cite{pumplin}. This contribution may be calculated perturbatively.
In order to compare the results with NLO QCD calculations, the contribution due to the underlying event must be subtracted from the jet cone.
CDF assumes an uncertainty of $30\%$ on the underlying event contribution to jets. This is the dominant source of systematic uncertainty for the CDF inclusive jet cross section at low E$_t$~\cite{jet_inc_sigma}. \par
In order to estimate the contribution of the underlying event in jet events, we consider two cones of radius $0.7$ at the same rapidity and at $\pm 90^\circ$ in azimuth from the most energetic jet in the events. The cone size is the same as used to reconstruct jets in the inclusive jet cross section analysis at CDF. We require the leading jet to be in the rapidity region $|\eta|<0.7$. 
The two cones are located in a semi-quiet region, far away from the two leading jets, but still in the central rapidity region. 
For each event we label the cone which has the maximum $\sum$p$_t$ ($max$ cone) and the cone with minimum $\sum$p$_t$ ($min$ cone), where $\sum$p$_t$ refers to the sum of the transverse momentum of the tracks inside the cone.  
The third parton, in NLO perturbative calculation of jet production, may fall in at most one of these two regions.  
In this approximation, the difference between the $max$ and the $min$ cone provides information on this contribution, while the $min$ cone  gives an  indication of the level of underlying event.  \par

\begin{figure}[t]
\epsfxsize=14.5pc 
\epsfbox{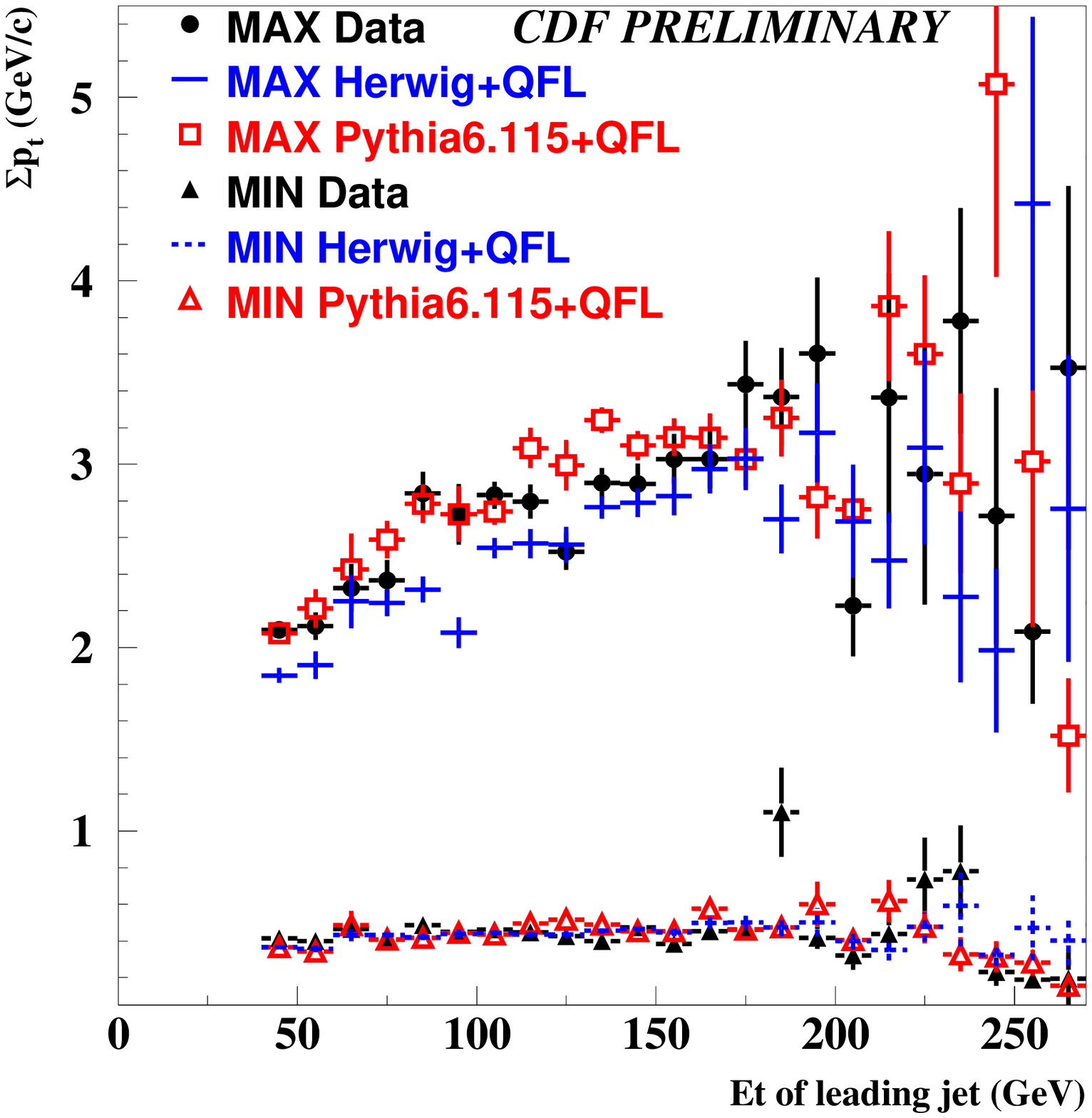} 
\epsfxsize=14.5pc 
\epsfbox{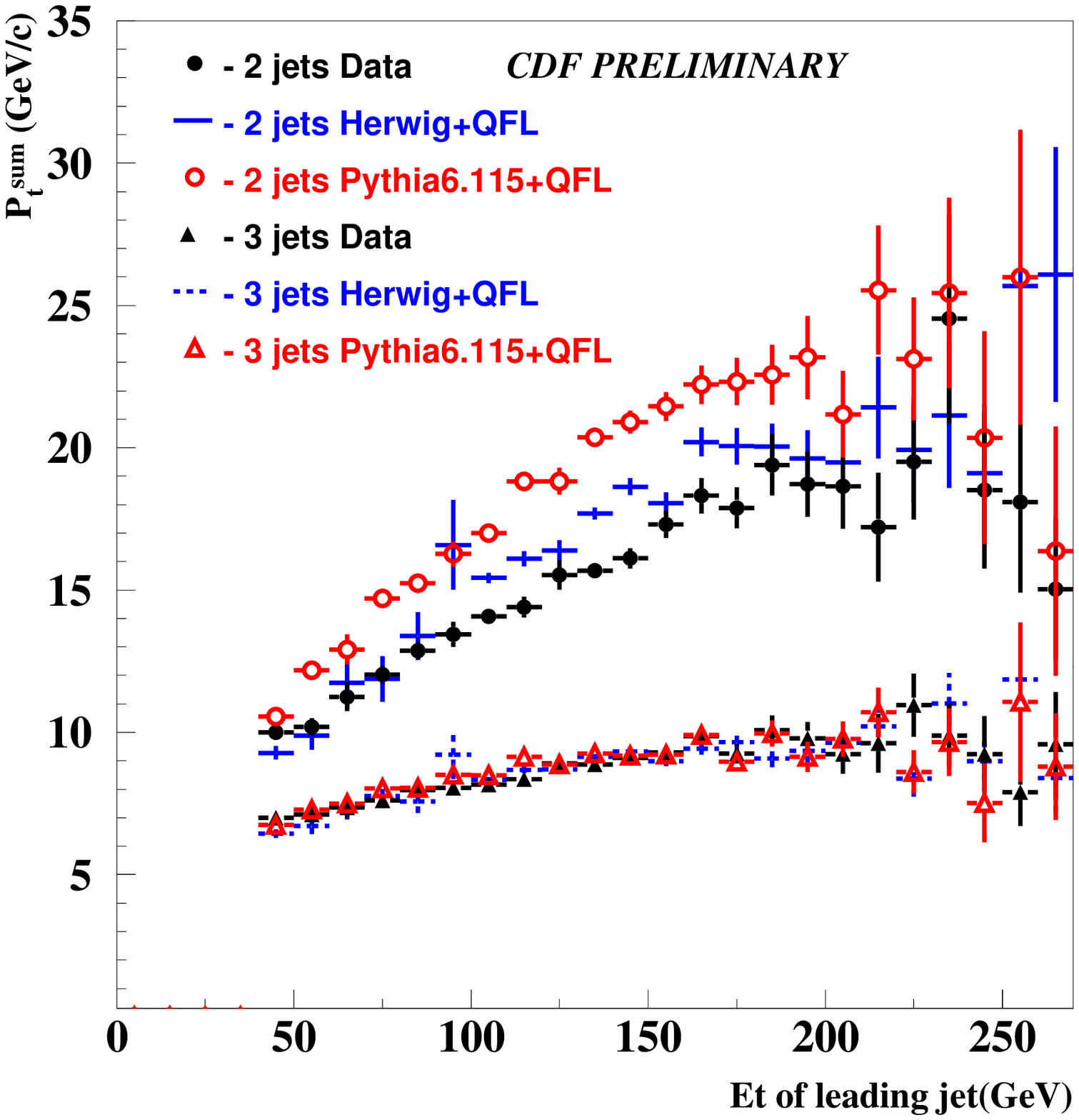} 
\caption{On the left: $\sum$p$_t$ inside the $max$ and $min$ cone in jet events at a center of mass energy of 1800 GeV as a function of E$_t$ of the leading jet. On the right: the \it{Swiss Cheese}\rm~ configuration is shown as a function of the E$_t$ of the leading jet~\label{pt_max_min_1800}}
\end{figure}
We use CDF data from Run1b, at a center of mass energy of 1800 GeV, and data collected in December 1995 from Run1c at $\sqrt{s}=630$ GeV.
In jet events at 1800 GeV, we consider four different data samples, each with E$_t$ of the leading jet larger than 40, 75, 100 and 130 GeV. We use two Monte Carlo programs, Pythia~\cite{pythia} and Herwig~\cite{herwig}, to generate four samples with the same cuts as the ones applied on data. The results from the Monte Carlo are then passed through QFL, the fast CDF detector simulation.  
We correct the tracks for the reconstruction efficiency and require p$_{t~ track} > 0.4$ GeV/c. Systematic uncertainties have been estimated to be of the order of $10\%$ or less.\par
Figure~\ref{pt_max_min_1800} (left) shows  $\sum$p$_t$ inside the $max$ and $min$ cone as a function of E$_t$ of the leading jet. We use Herwig with its default parameters for the soft physics description and tune  Pythia to attempt to reproduce the data. 
The regularization scale of the transverse momentum spectrum for multiple interactions (p$_{t0}$)~\cite{pythia_soft} has been set to 2 GeV/c.
Herwig+QFL, Pythia6.115+QFL and the data have a similar behavior for the $max$ and $min$ cone: the $min$ cone stays flat while the $max$ cone increases with the  E$_t$ of the leading jet.
The simulation reproduces the level of transverse momentum in the $min$ cone in CDF data. In the $max$ cone, Pythia is about 300 MeV larger than Herwig, with the data being in the middle. Since the $min$ cone is well reproduced, this difference is probably  due to the different algorithms used by the two Monte Carlo programs to generate parton shower.\par
An experimental approach to study the underlying event, complementary to the $max$ and $min$  cones in jet events, is the measurement of the sum of the transverse momentum in the central region ($|\eta| < 1$) excluding the transverse momentum of the particles inside a cone 0.7 from the center of the two or three most energetic jets in the event. 
This configuration is called \it{Swiss Cheese}\rm~ and is shown in figure~\ref{pt_max_min_1800} (right).
 There is a good agreement between data and simulation when the three most energetic jets in the event are subtracted. Here, the contribution from a third parton in the event should be subtracted. Again, when only two jets are subtracted, Pythia is larger than Herwig.\par
At $\sqrt{s} = 630$ GeV, the simulation agrees with data. Here, in order to reproduce the data, the regularization scale of the transverse momentum spectrum for multiple interactions (p$_{t0}$) in Pythia has been set to 1.4 GeV/c. A dependence of p$_{t0}$ on the center of mass energy has been implemented in versions of Pythia after 6.12. However, the dependence in Pythia is slower than what we found in this analysis and the default values underestimate the number of charged particles produced by  multiple interactions.

\section{Properties of soft minimum bias events}

\begin{figure}[t]
\epsfxsize=14pc 
\epsfbox{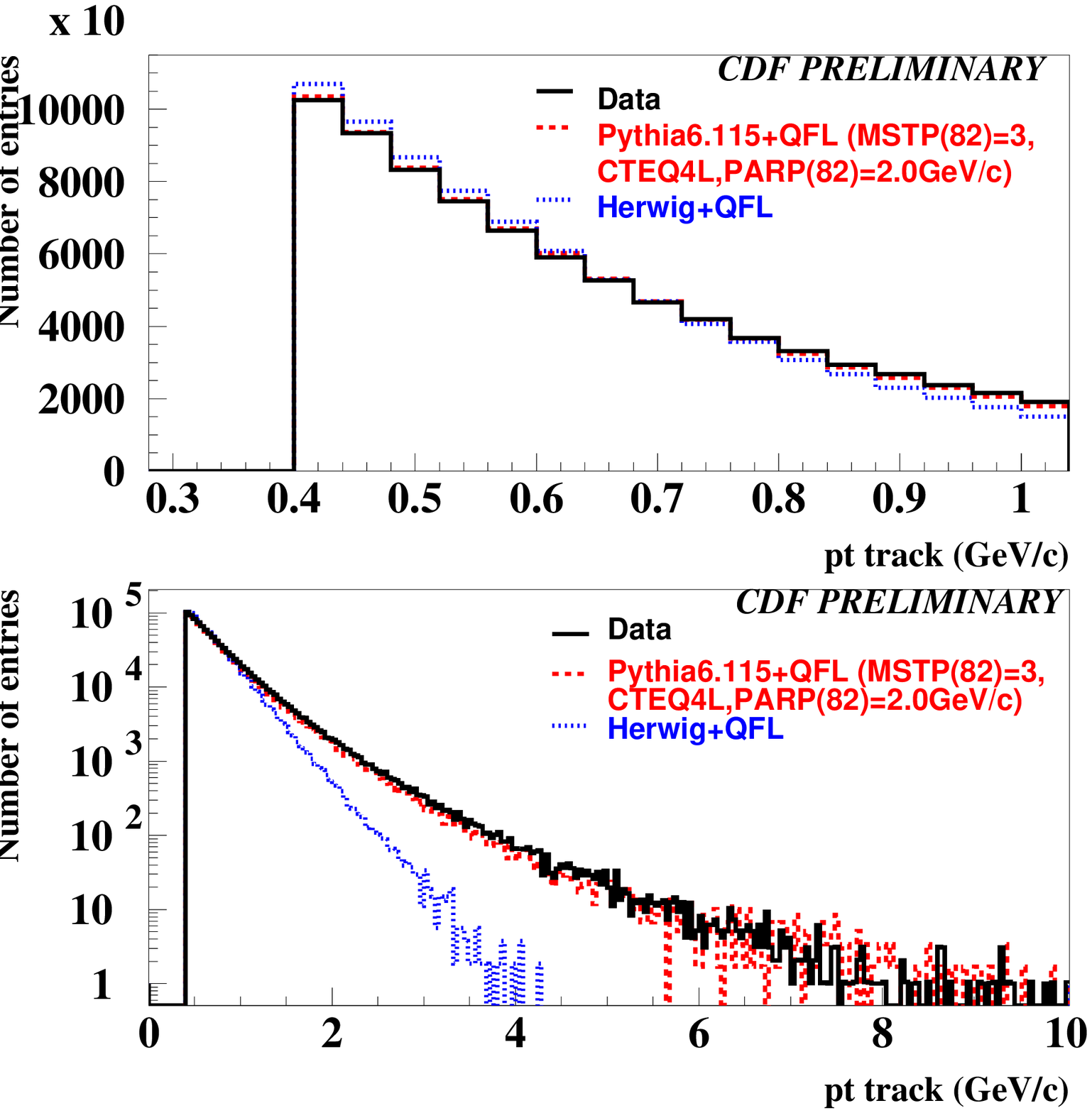} 
\epsfxsize=13pc 
\epsfbox{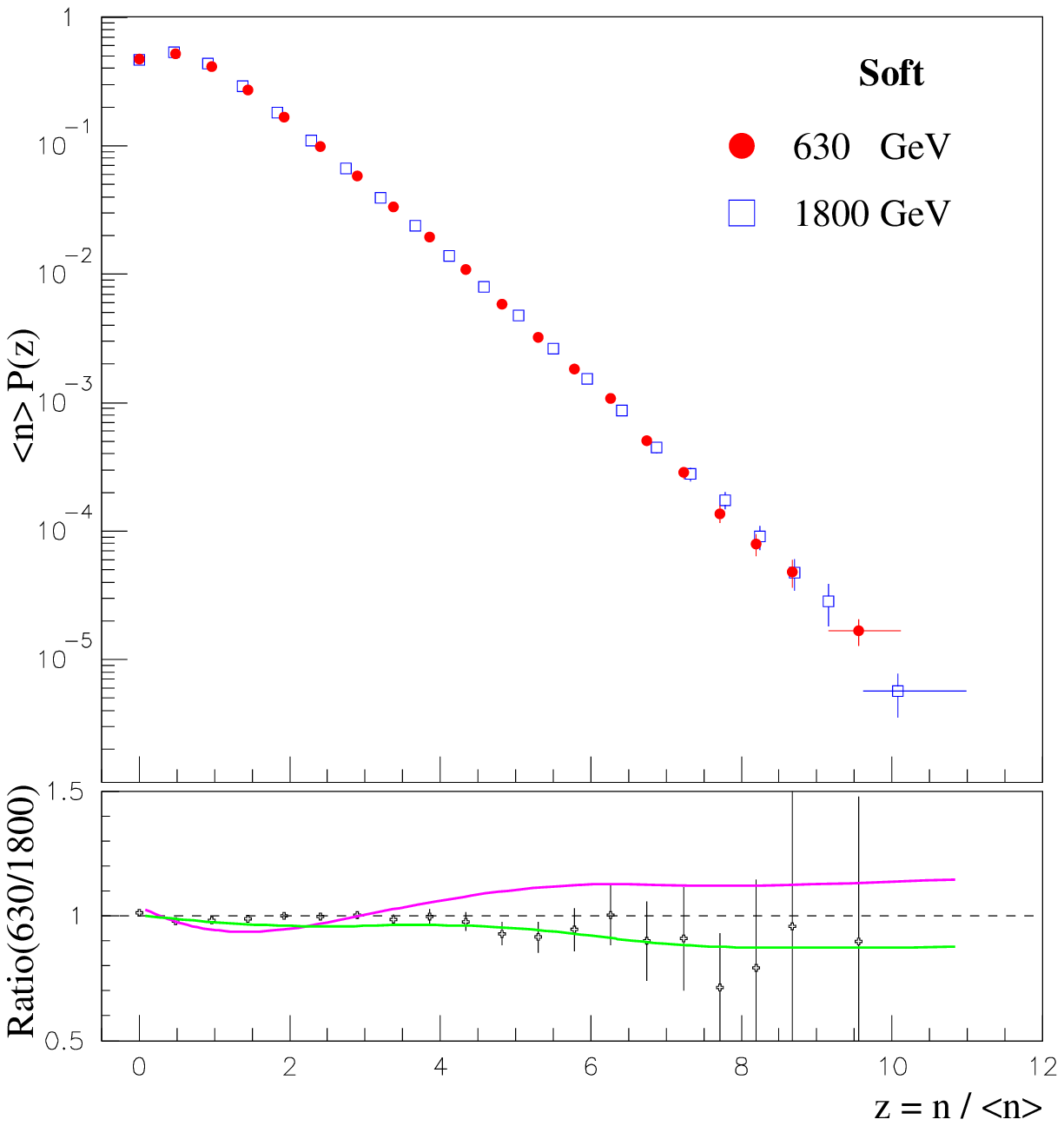} 
\caption{On the left: p$_t$ of the tracks in the central rapidity region ($|\eta<0.7$) at $\sqrt{s}=1800$ GeV in minimum bias events. The number of entries in the simulation has been normalized to the number of entries in the data. On the top part of the figure, only the region with p$_t<1$ GeV is shown in linear scale. On the right: multiplicity in the central rapidity region ($|\eta|<1$) in KNO variables. On the bottom of the plot the ratio of the the two distribution is shown. The two continuous lines delimit the band of all systematic uncertainties 
 \label{mb_1800}}
\end{figure}
In CDF, the minimum bias trigger requires coincident hits in scintillator counters located at 5.8 m from the interaction point. It samples double diffractive and non diffractive interactions from beam-beam collisions. 
The diffractive contribution to  overall cross section is small. \par
Minimum bias events have been analyzed at both 1800 and 630 GeV. Figure~\ref{mb_1800}~ (left) shows the transverse momentum distribution of the tracks in the central region ($|\eta|<0.7$) at $\sqrt{s}=1800$ GeV. 
Pythia reproduces the CDF data well, while Herwig does not generate tracks with p$_t > 5$ GeV, indicating a lack of semi-hard physics description in its model for the minimum bias events. At $\sqrt{s}=630$ GeV, neither Herwig or the tuned Pythia reproduce the data correctly.\par
We split the minimum bias sample in two subsamples: a \it{soft}\rm~ one, with events without clusters of E$_t > 1.1$ GeV in the rapidity region $|\eta|< 2.4$, and a \it{hard}\rm~ one,  with events containing  at least one of such clusters.
In figure~\ref{mb_1800} (right), the multiplicity in the central rapidity region ($|\eta|<1$), defined as the number of selected tracks in the event, is shown, for the \it{soft}\rm~ samples at 1800 and 630 GeV, in KNO variables. The distribution shows the validity of KNO~\cite{KNO} scaling in \it{soft}\rm~ minimum bias events, while the \it{hard}\rm~ sample slightly violates the KNO scaling.

\section{Conclusions}

An improved tuning of Monte Carlo programs and understanding of the underlying event and minimum bias events at the Tevatron, is also desired in order to study the complex physics environment at the LHC, where about 25 minimum bias events are superimposed to any event of interest.


\end{document}